\documentclass[a4paper]{jpconf}
\usepackage{graphicx}
\usepackage{amsmath}
\begin{document}
\title{The discrete charm of flavour and CP violation}

\author{\underline{G C Branco} and M N Rebelo}

\address{Centro de F\'isica Te\'orica de Part\'iculas -- CFTP and Dept de F\' \i sica
Instituto Superior T\'ecnico -- IST, Universidade de Lisboa (UL), Av. Rovisco Pais,
P-1049-001 Lisboa, Portugal}

\ead{gbranco@tecnico.ulisboa.pt and rebelo@tecnico.ulisboa.pt}

\begin{abstract}
We point out that in the Standard Model (SM) there is no explanation why 
$|V_{23}|^2 + |V_{13}|^2$ is of order $10^{-3}$. 
A framework is described for explaining this small
mixing, involving the introduction of vector-like quarks. A symmetry is 
introduced so that at a first stage  $V_{CKM} = 1\>\!\!\!\mathrm{I}$
and only the third generation 
of quarks acquires a mass.
It is shown that when interactions of vector-like quarks are taken 
into account a realistic 
quark mass spectrum is generated together with a correct $V_{CKM}$ matrix. 
\end{abstract}

\section{Introduction}
The closely related  questions of Flavour and CP Violation are two of the major 
open problems in Particle Physics.
Discrete symmetries have played a major role in the development of Particle Physics. 
The symmetries C, P, CP
and T are all violated in Nature. Yet, they are very important, since they are 
conserved by electromagnetic 
and strong interactions. Only CPT is not violated, as far as we know today. 
Parity is explicitly violated in the Standard 
Model (SM), but in the SM 
CP can be violated explicitly in the Yukawa sector provided that there are
three or more fermion generations. Spontaneous CP violation, 
can only occur if the scalar sector of the SM is extended. The first model
of spontaneous CP violation was proposed by Lee \cite{Lee:1973iz} in the context 
of the SM extended with one additional Higgs doublet, in order to
accommodate the experimental evidence for CP violation in the quark sector 
at a time when it was not yet known that there were more than 
two generations of quarks. Models with two Higgs doublets have a very rich
phenomenology \cite{Gunion:1989we,Branco:2011iw}
and they provide very interesting candidates for physics beyond the SM,
which have the potential of being tested at the LHC. 
There is, by now, clear experimental evidence that the Cabibbo-Kobayashi-Maskawa 
(CKM) matrix is complex \cite{Botella:2005fc}.
Therefore, in any model of spontaneous CP violation, with the breaking of CP arising 
from a CP violating vacuum, it must be possible to generate a complex
CKM matrix from the vacuum phase, in order to conform to experiment. This makes it non-trivial 
to have viable models of spontaneous CP violation. The simplest realistic  model 
of spontaneous CP violation, involves the
addition to the SM of a vector-like isosinglet quark and a complex 
singlet scalar \cite{Bento:1990wv,Bento:1991ez,Branco:2003rt}. Models with 
several Higgs doublets have potentially large Higgs mediated
flavour changing neutral currents. 
Experimental data show that the latter are extremely suppressed thus implying
the need for a mechanism to avoid, in a natural way, sizeable effects. One 
option is to use a symmetry that eliminates these currents at tree level, 
this is the natural flavour conservation paradigm \cite{Glashow:1976nt, Paschos:1976ay}.  
In alternative it is possible to use a symmetry that leads to tree level 
effects suppressed by  small entries of $V_{CKM}$ \cite{Branco:1996bq}. 
These are the so-called BGL models which have been extended later-on 
\cite{Botella:2009pq, Botella:2011ne}
and where the suppression conforms with the 
experimental evidence.\\

At present, all experimental results on quark mixing and CP violation in the quark 
sector, are in essential agreement 
with experiment. However, there is plenty of room for New Physics. 
CP violation has profound implications for cosmology, since CP breaking is one of 
the required ingredients \cite{Sakharov:1967dj}  in order 
to generate the observed Baryon Asymmetry of the Universe (BAU). 
By now, it is clear that in the SM it is not possible
to generate sufficient BAU. One of the reasons for this is the smallness of 
CP violation in the SM. 
This is a clear indication that there should be additional sources of CP violation, 
beyond the one present in the SM.
The great challenge is to find out where are the new sources of CP violation and how 
they could be detected. In the quark sector, an interesting possibility may arise 
if vector-like quarks are added to the spectrum of the SM 
\cite{Bento:1991ez,delAguila:1985ne,Branco:1986my,
Lavoura:1992qd,Branco:1992wr, Barenboim:1997qx,Barenboim:2000zz,Barenboim:2001fd,
AguilarSaavedra:2002kr,AguilarSaavedra:2004mt,Botella:2008qm,Higuchi:2009dp,
Botella:2012ju,Fajfer:2013wca,Aguilar-Saavedra:2013qpa,
Ellis:2014dza,Alok:2015iha,Ishiwata:2015cga, Bobeth:2016llm}.
These quarks 
can have bare masses in the Lagrangian, since their mass terms are invariant 
under the gauge symmetry. They may also
receive their mass from the couplings to SU(2) singlet scalars, once these acquire a 
vacuum expectation value (vev). In any case, they mix
with the standard quarks, and the $3 \times 3$ unitarity of the CKM 
matrix can be violated. 
These violations of $3 \times 3$ unitarity
lead in general to Z-mediated Flavour-Changing-Neutral-Currents (FCNC) at 
tree level, which are naturally 
suppressed. In spite of this suppression, these FCNC,
like $B_d-\overline B_d$ mixing, $B_s-\overline B_s$ mixing and 
$D^0-\bar D^0$ mixing \cite{Branco:1995us},
can contribute to 
processes which in the SM only occur at loop level. \\

Another possible source of CP breaking is leptonic CP violation. It is well known 
that in the SM neutrinos are strictly 
massless, so the discovery of neutrino oscillations, implying that at least 
two of the neutrinos have mass, is
a clear experimental evidence of Physics Beyond the SM. Let us assume that one 
adds three right-handed neutrinos 
to the SM. Then the most general renormalizable Lagrangian must include 
Majorana-type bare mass terms for the right-handed neutrinos, together 
with Dirac neutrino masses,
leading to the seesaw mechanism
\cite{Minkowski:1977sc,Yanagida,Glashow,GellMann:1980vs,Mohapatra:1979ia}
of type one, providing an elegant explanation for the 
smallness of the neutrino masses. In this scenario, there are three types of 
CP violation which may arise: CP violation
at low energies of Dirac type, CP violation at low energies of Majorana type and 
CP violation at high energy which
provides a very attractive scenario of Baryogenesis through 
Leptogenesis \cite{Fukugita:1986hr}. 
At low energies one may detect leptonic 
CP violation of Dirac type through neutrino oscillations,
which is one of the major tasks for future experiments. 
Majorana CP phases 
are difficult to detect at low energies 
but they affect double-beta decay. An important question is whether 
there is a relationship between 
leptonic CP violation at low energy and CP violation entering in Leptogenesis. 
It has been shown that in general 
such a relationship does not exist \cite{Branco:2001pq,Rebelo:2002wj}
unless one introduces leptonic flavour symmetries.

\section{Comparing the Quark and Leptonic Sectors}

In the SM, the flavour structure of Yukawa couplings is not constrained by the
gauge symmetry which implies that
the Yukawa matrices are $3 \times 3$ general complex matrices. This leads to the existence 
of 36 parameters in the 
quark flavour sector. It is clear that there is a large redundancy, since in the 
quark flavour sector we have only ten physical parameters, namely the six quark masses 
plus the four physical parameters of the CKM matrix. In the lepton sector, 
one encounters an entirely analogous situation in the case of massive Dirac neutrinos. 
In the case of Majorana 
neutrinos, one has two extra CP violating phases which reflect the Majorana 
character of neutrinos.
This large flavour redundancy  is due to the freedom that one has to make weak 
basis (WB) transformations which, for the quark sector, can be written: 
\begin{equation}
d_L^{0} \rightarrow U^\prime d_L^{0}, \quad u_L^0 \rightarrow U^\prime u_L^0,
\quad d_R^{0} \rightarrow  V^\prime d_R^{0} \quad u_R^0 \rightarrow  W^\prime u_R^0
\end{equation}

As previously mentioned, the Yukawa matrices are arbitrary 
complex matrices. However, one can make a weak 
basis transformation so that both $Y_u$, $Y_d$  become Hermitian
matrices. In the context of specific anz\" atze with texture zeros it is 
common to consider Hermitian matrices (see, for example, \cite{Ramond:1993kv})
yet, this need not be the case \cite{Branco:1994jx}.
The large redundancy in Yukawa couplings
renders it specially difficult to extract from the data some flavour symmetry, 
in a bottom up approach. Even if such symmetry exists, in what WB would the 
symmetry be apparent? There is a specially convenient WB where
one of the quark mass matrices ($M_d$ or $M_u$) is diagonal real and the 
other matrix is Hermitian. It can be readily seen that in this basis the number 
of free parameters in the quark mass matrices equals the number of physical 
measurable parameters, namely the six quark masses and the four physical parameters 
in $V_{CKM}$.

The discovery of neutrino oscillations providing experimental evidence for 
non-vanishing neutrino masses, brought 
further intriguing aspects to the Flavour Problem. Why are the pattern of 
fermion mixings so different in the quark
and lepton sectors?  Do we understand why lepton mixing is large? Do we understand 
why quark mixing is small?
An important difference between the quark and lepton sectors has to do with 
experiment. In the quark sector 
the $V_{CKM}$ matrix is overdetermined, in the sense that once we measure 
$V_{us}$, $V_{ub}$ and  $V_{cb}$ through strange particle and
B-meson decays, there is a large amount of physical quantities like 
$B_d-\overline B_d$, $B_s-\overline B_s$ mixings, the strength of CP 
violation in the kaon sector as well as CP violation in the B sector, 
with the measurement of the angles $\beta$  and 
$\gamma$, which within the SM, have to be fitted through a single parameter, 
namely the CP violating phase entering in the parametrization of $V_{CKM}$. 
This is to be contrasted to the situation encountered in the lepton  sector. 

Let us assume that neutrinos are Majorana particles and consider the effective 
neutrino mass matrix at low energies. The leptonic mass terms are then of the form:
\begin{equation}
{\mathcal L}_{\mbox{mass}} = 
-\frac{1}{2}  {\nu_L^{0}}^T C^{-1} m_\nu \nu_L^{0} 
- \overline{\ell_L^0 } m_\ell \ell_R^0 
+ \text{h.c.}\ .
\label{mb}
\end{equation}
The neutrino mass matrix is a complex symmetric matrix, with six real parameters 
and three phases. If we work in the WB where the charged lepton 
mass matrix is diagonal and real, the number of independent parameters of the
neutrino mass matrix is equal to the number of physical quantities derived from it,
to wit, three neutrino masses, and six parameters -
three mixing angles and three CP violating phases -
appearing in the leptonic mixing matrix. It has been pointed out that it is not possible
to completely determine the neutrino mass matrix from viable experimental 
measurements \cite{Frampton:2002yf}.
This motivated the authors of Ref.~\cite{Frampton:2002yf} to
consider a restricted class of neutrino mass matrices with zero textures 
consistent with all experimental data and to explore their phenomenological 
implications. The imposition of these textures is based on the underlying 
idea that they are linked to a flavour symmetry. 
The effective neutrino mass matrix written in
Eq.~(\ref{mb}) must be generated through physics beyond the SM. The seesaw  
framework is a simple and elegant possibility. The authors of Ref.~\cite{Kageyama:2002zw}
studied seesaw realisations of the texture zeros proposed in \cite{Frampton:2002yf}.
Implications for CP violation
in the leptonic sector at high energies, and in particular 
for leptogenesis, of these seesaw realizations were analysed in \cite{Kaneko:2002yp}.
With two right-handed neutrinos only, zero textures allow to 
establish a connecting link between the sign of the baryon 
asymmetry of the universe and the sign of the CP violating
parameter which will be measured in low energy neutrino oscillations 
 \cite{Frampton:2002qc}. With three right-handed neutrinos
these signs cease to be correlated \cite{Branco:2002xf} but a connection still exists.
In this respect the Casas and Ibarra parametrisation \cite{Casas:2001sr}
is very useful to show that
zeros in the Dirac type neutrino mass matrix, in the basis where the
charged lepton mass matrix is diagonal, lead to orthogonality 
relations such that for some of the textures the variables relevant for 
leptogenesis can be fully determined in terms of low energy parameters and 
heavy neutrino masses \cite{Branco:2005jr,Branco:2007nb,Bjorkeroth:2015tsa}.
An alternative way of reducing the number of free parameters in the 
Lagrangian is by means of discrete symmetries, such as for example 
an $A_4$ symmetry \cite{Hagedorn:2009jy,Branco:2009by} (for a review see \cite{Altarelli:2010gt}), 
leading, once again, to definite predictions.
In general without a flavour model
the three CP violating phases appearing at low energies and the three phases
relevant for high energy leptonic CP violation, and in particular Leptogenesis, 
are not related \cite{Branco:2001pq,Rebelo:2002wj}. \\

\section{The Origin of CP Violation}

In  the SM with three or more generations, CP violation may originate at 
the Lagrangian level with complex 
Yukawa couplings, as it has been pointed out by Kobayashi and Maskawa 
\cite{Kobayashi:1973fv}. 
The most elegant way to show this 
consists of separating the Lagrangian of the SM in two parts:
\begin{equation}
{\cal{L}}= {\cal{L}}_{CP} + {\cal{L}}_{\mbox{remaining}} 
\end{equation}
${\cal{L}}_{CP}$ consists of the gauge part of the Lagrangian which necessarily 
conserves CP. One can then consider the most
general CP transformation which leaves ${\cal{L}}_{CP}$ invariant. Applying this CP 
transformation to ${\cal{L}}_{\mbox{remaining}}$ one can easily 
derive that the necessary condition for CP invariance for any number of 
fermion generations is \cite{Bernabeu:1986fc}:
\begin{equation}
\mbox{Tr} \left[  H_u, H_d \right]^3  = 0 
\label{trcom}
\end{equation}
The condition of Eq.~(\ref{trcom}) is automatically satisfied for one or two generations. For 
three generations Eq.~(\ref{trcom}) is a necessary and 
sufficient condition for CP invariance and it can be expressed in 
terms of quark masses and mixing:
\begin{eqnarray}
\mbox{tr} \left[  H_u, H_d \right]^3  = 6i & (m^2_t - m^2_c) &   
(m^2_t - m^2_u) \quad  (m^2_c - m^2_u) \times \nonumber \\ 
& \times (m^2_b - m^2_s)  & (m^2_b - m^2_d) \quad (m^2_s - m^2_d) \quad \mbox{Im} Q_{uscb}
\label{eq31}
\end{eqnarray}
where $Q$ stands for a rephasing invariant quartet of $V_{CKM}$,
defined by $Q_{\alpha i \beta j} \equiv V_{\alpha i}  V_{\beta j}  V^\ast_{\alpha j}  
V^\ast_{\beta i} $ ($\alpha \neq \beta$, $ i \neq j$).

Another scenario for the generation of CP violation is the one put forward by 
Lee \cite{Lee:1973iz} about the same time that the KM mechanism 
\cite{Kobayashi:1973fv}   was proposed. The idea 
is that CP is a good symmetry of the Lagrangian but the vacuum is not CP 
invariant.

In general a realistic model of Spontaneous CP violation has to 
fulfil the following conditions.

i) The model should have a natural suppression mechanism for  
FCNC both in the scalar and vector sectors.

ii) The vacuum CP violating phase phase should be able to generate a 
complex CKM matrix. This requirement 
arises from the fact that experimentally there is clear evidence that 
the CKM matrix is non-trivially complex even 
if one allows for the presence of New Physics.

 The minimal viable model of spontaneous CP violation involves 
the following extension of the SM:

i) Introduction of at least one vector-like quark with charge either -1/3 or +2/3. 

ii) Introduction of one complex scalar field, singlet under the gauge group, 
which we denote S.

With the above mentioned addition to the SM, one achieves spontaneous CP violation 
through a non-trivial phase in the vev of the scalar isosinglet S. For definiteness, 
let us assume that we have introduced a down type  vector-like 
quark. This leads to a $4 \times 4$  quark mass matrix in the down sector and the quark 
mixing matrix becomes a
$3 \times 4$ matrix, leading to deviations of $3 \times 3$ unitarity in the 
CKM matrix connecting standard quarks and also to
Z-mediated FCNC in the down sector. One of the nice features of this class of 
models is the fact that both deviations 
of $3 \times 3$ unitarity and tree level FCNC are naturally suppressed  by 
the ratio of masses $m^2/M^2$ of standard quarks and 
of the heavy isosinglet quark. Recall that since the mass term of the 
vector-like quark is gauge invariant, its value can
naturally be much larger than the electroweak scale. There is mixing between 
standard quarks and the vector-like 
quark which in turn generates at low energies a complex $3 \times 3$ effective 
down quark mixing. Upon diagonalisation of the quark mass matrices,  this 
leads to a complex $3 \times 3$ CKM matrix, in agreement with experiment.  
As mentioned  above there are naturally suppressed FCNC at tree level which 
in spite of this suppression can contribute
to $B_d-\overline B_d$ and $B_s-\overline B_s$ mixings.

\section{Generating a Viable Model of Quark Mixing with Vector-Like Quarks}

In the SM the Yukawa couplings $Y_u$, $Y_d$, responsible for  generating the up and 
down quark mass matrices, are
two independent complex matrices. As a result, we show next that there is
no reason for having the CKM matrix close
to the identity, even if one takes into account the strong hierarchy of 
quark masses. In order to prove this result we shall 
consider the extreme chiral limit, where only the top and the bottom quarks have a mass.

\subsection{The Extreme Chiral Limit}

In the EC limit the up and down quark mass matrices are two rank one 
matrices which can be written as :
\begin{equation}
M_{d}={U_{L}^{d}}^{\dagger }\ \mbox{diag}(0,0,m_{b})\ {U_{R}^{d}},\qquad
M_{u}={U_{L}^{u}}^{\dagger }\ \mbox{diag}(0,0,m_{t})\ {U_{R}^{u}}
\label{loo}
\end{equation}                                                                                                                                                                  
Note that no generality is lost in writing in Eq.~(\ref{loo}) the quark 
masses in a  given ordering, taking into account 
that a permutation which would change these positions could always be 
incorporated in the matrices $U_{L,R}^{d,u}$ .

Keeping in mind that at this stage the first two generations are massless, 
one can make an arbitrary redefinition of 
the massless quarks through a unitary transformation with the following form:
\begin{equation}
W_{u,d}=\left[ 
\begin{array}{cc}
X_{u,d} & 0 \\ 
0 & 1%
\end{array}
\right]   \label{lii}
\end{equation}
where $X_{u,d}$ are $2\times 2$ unitary matrices. One can then show that the 
quark mixing  matrix  $V_{CKM}$ is an orthogonal
matrix mixing only the third and second quark generation, parametrised by an angle 
$\alpha$ which is entirely
arbitrary in the EC limit. It is clear that in the SM the smallness 
of $|V_{23}|^2 + |V_{13}|^2$ has no relation
with the fact that the quark mass ratios are small. The experimentally observed 
smallness of $|V_{23}|^2 + |V_{13}|^2$ may be an indication of the need of 
a flavour symmetry.

\section{Obtaining Small Mixing Through a Symmetry}

Let us introduce the following $Z_6$ flavour symmetry in the SM:
\begin{align}
\label{SSM}
Q_{L1}^{0}& \rightarrow e^{i\tau }\ Q_{L1}^{0} &\quad Q_{L2}^{0}&\rightarrow
e^{-2i\tau }\ Q_{L2}^{0} &\quad Q_{L3}^{0}&\rightarrow e^{-i\tau }\ Q_{L3}^{0} \nonumber \\ 
\nonumber \\ 
d_{R1}^{0}&\rightarrow e^{-i\tau }d_{R1}^{0} &\quad d_{R2}^{0}&\rightarrow
e^{-i\tau }d_{R2}^{0} &\quad d_{R3}^{0}&\rightarrow e^{-2i\tau }d_{R3}^{0}  \\
\nonumber \\ 
u_{R1}^{0}&\rightarrow e^{i\tau }u_{R1}^{0} &\quad u_{R2}^{0}&\rightarrow
e^{i\tau }u_{R3}^{0} &\quad u_{R3}^{0}&\rightarrow u_{R3}^{0};\ &\quad 
\Phi &\rightarrow e^{i\tau }\Phi; &\quad \tau & = \frac{2 \pi}{6} \nonumber
\end{align}
where the $Q_{Lj}^{0}$  stand for the left-handed quark doublets, $d_{Rj}^{0}$  and 
$u_{Rj}^{0}$ denote right-handed quark singlets 
and $\Phi $ is the scalar doublet. The Yukawa interactions can be written:
\begin{equation}
\mathcal{L}_{\mathrm{Y}}=\left[ -{\overline{Q}_{Li}^{0}}\ \,\Phi \,Y_{d}\,\
d_{Rj}^{0}-\,{\overline{Q}_{Li}^{0}}\ \tilde{\Phi}\,Y_{u}\,\ u_{Rj}^{0}
\right] +\mbox{h.c.},
\end{equation}
As a result of the $Z_6$ symmetry, the Yukawa couplings have the following flavour structure:
\begin{equation}
Y_{d}=\left[ 
\begin{array}{ccc}
0 & 0 & 0 \\ 
0 & 0 & 0 \\ 
0 & 0 & \times
\end{array}
\right] ,\qquad Y_{u}=\left[ 
\begin{array}{ccc}
0 & 0 & 0 \\ 
0 & 0 & 0 \\ 
0 & 0 & \times
\end{array}
\right]  \label{ydyu}
\end{equation}
which leads to $V_{CKM}= 1\>\!\!\!\mathrm{I}$, at this stage.

\subsection{Generation of Quark Masses for the First Two Families}

We show next that one may generate masses for the first two generations, 
through the introduction of
vector-like quarks. For definiteness, we introduce three down 
($D^0_{Li}$, $D^0_{Ri}$,)and three up ($U^0_{Li}$, 
$U^0_{Ri}$,)  vector-like
quarks which are singlets under  the gauge group. The Yukawa interactions can now be written:
\begin{equation}
\mathcal{L}_{\mathrm{Y}}=\left[- {\overline Q_{Li}^{0}} \,\Phi\, (Y_d)_{i
\alpha} \, d^0_{R\alpha} - \, {\overline Q_{Li}^{0}}\tilde\Phi \, (Y_u)_{i
\beta} \, u^0_{R\beta} \right] + \mbox{h.c.},  \label{YYY}
\end{equation}
where the index i runs from 1 to 3 while the indices $\alpha$ and $\beta$ 
run from 1 to 6. We introduce the following generic bare mass terms
\begin{equation}
\mathcal{L}_{b.m.} = [- {\ \overline D^0_{Lj}} (\eta_d)_{j \alpha} \,
d^0_{R\alpha} - {\overline U^0_{Lk}} (\eta_u)_{k \beta} \, u^0_{R\beta} ] + 
\mbox{h.c.}  \label{bbb}
\end{equation}
The $Z_6$ symmetry is extended to the full Lagrangian, and we further 
introduce a complex scalar singlet denoted $S$. Soft-breaking terms are 
introduced in the scalar sector to avoid a massless Goldstone boson
and also to solve the domain-wall problem. The couplings of the scalar 
singlet to the quark singlets can be written: 
\begin{equation}
\mathcal{L}_{\mathrm{g}} = [- {\ \overline D^0_{Lj}} [({g_d})_{j \alpha} S +
({g_d^\prime})_{j \alpha} S^\ast ] \, d^0_{R\alpha} - {\overline U^0_{Lk}} [(
{g_u})_{k \beta} S + ({g_u^\prime})_{k \beta} S^\ast ] \, u^0_{R\beta} ] + 
\mbox{h.c.}  \label{ggg}
\end{equation}
After gauge symmetry breaking one obtains:
\begin{equation}
\mathcal{L}_{\mathrm{M}} = \left[- \frac{v}{\sqrt{2}}\, {\ \overline d^0_{Li}
} (Y_d)_{i \alpha} \, d^0_{R\alpha} - \frac{v}{\sqrt{2}} {\overline u^0_{Li}}
(Y_u)_{i \alpha} \, u^0_{R\alpha} - {\overline D^0_{Li}} (\mu_d)_{i \alpha}
\, d^0_{R\alpha} - {\overline U^0_{Li}} (\mu_u)_{i \alpha} \, u^0_{R\alpha} 
\right] + \mbox{h.c.}
\end{equation}
In a more compact form one can write:
\begin{equation}
\mathcal{L}_{\mathrm{M}} = - \left( {\ \overline d^0_{L}} \ {\overline
D^0_{L}} \right) \mathcal{M}_d \, \left( 
\begin{array}{c}
d^0_{R} \\ 
D^0_{R}
\end{array}
\right) - \left( {\ \overline u^0_{L}} \ {\overline U^0_{L}} \right) 
\mathcal{M}_u \, \left( 
\begin{array}{c}
u^0_{R} \\ 
U^0_{R}
\end{array}
\right)
\end{equation}
where $ \mathcal{M}_d$,  $ \mathcal{M}_u$ are $6 \times 6$ matrices
\begin{equation}
\mathcal{M}_d = \left( 
\begin{array}{cc}
m_d & \omega_d \\ 
X_d & M_d
\end{array}
\right) \quad \mathcal{M}_u = \left( 
\begin{array}{cc}
m_u & \omega_u \\ 
X_u & M_u
\end{array}
\right)  \label{not}
\end{equation}
These mass matrices can be diagonalised through unitary transformations:
\begin{equation}
\begin{array}{lll}
\left( 
\begin{array}{c}
d_{L}^{0} \\ 
D_{L}^{0}
\end{array}
\right) =\left( 
\begin{array}{c}
A_{dL} \\ 
B_{dL}
\end{array}
\right) \left( 
\begin{array}{c}
d_{L}
\end{array}
\right) \equiv \mathcal{U}_{L}^{d}d_{L} &  & \left( 
\begin{array}{c}
u_{L}^{0} \\ 
U_{L}^{0}
\end{array}
\right) =\left( 
\begin{array}{c}
A_{uL} \\ 
B_{uL}
\end{array}
\right) \left( 
\begin{array}{c}
u_{L}
\end{array}
\right) \equiv \mathcal{U}_{L}^{u}u_{L} \\ 
&  &  \\ 
\left( 
\begin{array}{c}
d_{R}^{0} \\ 
D_{R}^{0}
\end{array}
\right) \equiv \mathcal{U}_{R}^{d}d_{R} &  & \left( 
\begin{array}{c}
u_{R}^{0} \\ 
U_{R}^{0}
\end{array}
\right) \equiv \mathcal{U}_{R}^{u}u_{R}
\end{array}
\end{equation}
One can then write the charged currents as:
\begin{equation}
\mathcal{L}_{\mathrm{W}}=-\frac{g}{\sqrt{2}}\left( {\overline{u}_{L}^{0}}
\gamma ^{\mu }d_{L}^{0}\right) \mathbf{W}_{\mu }^{+}+\mbox{h.c.}=-\frac{g}{
\sqrt{2}}\left( {\overline{u}_{L}}V\gamma ^{\mu }d_{L}\right) \mathbf{W}
_{\mu }^{+}+\mbox{h.c.}
\end{equation}
with $V=A_{uL}^{\dagger }A_{dL}$. The neutral gauge couplings are:
\begin{equation}
\mathcal{L}_{\mathrm{Z}}=\frac{g}{\cos {\theta _{W}}}Z_{\mu }\left[ \frac{1}{
2}\left( {\overline{u}_{L}}W^{u}\gamma ^{\mu }u_{L}-{\overline{d}_{L}}
W^{d}\gamma ^{\mu }u_{L}\right) -\sin ^{2}\theta _{W}\left( \frac{2}{3}
\overline{u}\gamma ^{\mu }u-\frac{1}{3}\overline{d}\gamma ^{\mu }d\right) 
\right] 
\end{equation}
with $W^{d}=V^{\dagger }V$ and $W^{u}=VV^{\dagger }$. In the quark mass
eigenstate basis the SM-like Higgs couplings, and the couplings of the would-be
Goldstone bosons $G^+$, $G^0$ can be written:
\begin{equation}
\begin{array}{l}
\mathcal{L}_{\mathrm{Mh}}=
-\frac{\sqrt{2}G^{+}}{v}\left[ \overline{u}_{L}V
\mathcal{D}_{d}d_{R}-\overline{u}_{R}\mathcal{D}_{u}Vd_{L}\right]  \\ 
\\ 
-i\frac{G^{0}}{v}\left[ \overline{d}_{L}W^{d}\mathcal{D}_{d}d_{R}-\overline{
u}_{L}W^{u}\mathcal{D}_{u}u_{R}\right] -\frac{h}{v}\left[ \overline{d}
_{L}W^{d}\mathcal{D}_{d}d_{R}+\overline{u}_{L}W^{u}\mathcal{D}_{u}u_{R}
\right] +\mbox{h.c.}
\end{array}
\end{equation}
The vector-like quarks transform in the following way under $Z_6$:
\begin{align}
D_{L1}^{0} & \rightarrow e^{-3i\tau }\ D_{L1}^{0}  &\quad 
D_{L2}^{0} & \rightarrow
e^{-2i\tau }\ D_{L2}^{0} &\quad  D_{L3}^{0}& \rightarrow e^{-i\tau }\ D_{L3}^{0}
\nonumber \\
\nonumber \\
D_{R1}^{0} & \rightarrow e^{-2i\tau }\ D_{R1}^{0} &\quad  D_{R2}^{0}& \rightarrow
e^{-3i\tau }\ D_{R2}^{0} &\quad  D_{R3}^{0}& \rightarrow D_{R3}^{0} \nonumber \\ 
  \\ 
U_{L1}^{0} & \rightarrow e^{-i\tau }\ U_{L1}^{0} &\quad  U_{L2}^{0}& \rightarrow
U_{L2}^{0} &\quad  U_{L3}^{0} & \rightarrow e^{i\tau }\ U_{L3}^{0} \nonumber \\ 
\nonumber  \\ 
U_{R1}^{0} & \rightarrow U_{R1}^{0} &\quad  U_{R2}^{0}& \rightarrow e^{-i\tau }\
U_{R2}^{0} &\quad  U_{R3}^{0} & \rightarrow e^{2i\tau }\ U_{R3}^{0};\ & \quad 
S\rightarrow e^{i\tau }\ S; & \quad \tau & = \frac{2 \pi}{6} \nonumber 
\end{align}
Diagonalisation of the quark mass matrices is performed through:
\begin{equation}
\mathcal{U}_{L}^{d \dagger} \mathcal{M}_{d}\ \mathcal{U}_{R}^d = \mathcal{D}
_d \equiv \mbox{diag} (d_d, D_d)  \label{plus}
\end{equation}
with:
\begin{equation}
\mathcal{U}_L = \left( 
\begin{array}{cc}
K & R \\ 
S & T
\end{array}
\right)
\end{equation}
Unitarity of $\mathcal{U}_L$ implies
$K K^\dagger = {1\>\!\!\!\mathrm{I}} - R R^\dagger$
as well as $K^\dagger K = {1\>\!\!\!\mathrm{I}} - S^\dagger S$.
Deviations of $3 \times 3$ unitarity are small since $R$ and $S$
are suppressed by the ratio $m/M$. 
The matrices $K_d$, $K_u$ 
can be computed from an effective Hermitian squared matrix given by:
\begin{equation}
\mathcal{H}_{eff} = (m m^\dagger + \omega \omega^\dagger) - (mX^\dagger +
\omega M^\dagger) (X X^\dagger + M M^\dagger)^{-1} ( X m^\dagger + M
\omega^\dagger)  \label{28}
\end{equation}
To an excellent approximation $K$ is the matrix that diagonalises 
$\mathcal{H}_{eff}$:
\begin{equation}
K^{-1} \mathcal{H}_{eff} K = d^2
\end{equation}
We assume that the $6 \times 6$ matrix ${\cal M}_d $ has a Froggatt-Nielsen
structure \cite{Froggatt:1978nt}: 
\begin{equation}
{\cal M}_d = \mu \left( 
\begin{array}{cccccc}
0  & 0 & 0 & 0 & 0 & 
\lambda^2 z  \\ 
0 & 0 & 0 & 0 & \lambda y  & 
0 \\ 
0 & 0 & 1 & \lambda x  & 0 & 
0 \\ 
0 & 0 & \frac{C_1}{\lambda} &  \frac{D_1}{\lambda^2} & 0
& 0 \\ 
\frac{A_2}{\lambda}  & \frac{B_2}{\lambda} & \frac{C_2}{\lambda} & 0 & 
\frac{D_2}{\lambda^2} & 0 \\ 
\frac{A_3}{\lambda} & 0 & \frac{C_3}{\lambda}& 0 & 
0 & \frac{D_3}{\lambda^2}
\end{array}
\right) ,  \label{calMd}
\end{equation}
with $\mu \approx  m_b$. This leads to the following structure for $\mathcal{H}_{eff}$:
\begin{equation}
\mathcal{H}_{eff} \sim \left( 
\begin{array}{ccc}
\lambda^6 r z^2 & \lambda^5 uyz & - \lambda^3 c_3 z \\  
 \lambda^5 u^* yz &  \lambda^4 r^\prime y^2 &  - \lambda^2 c_2 y \\
 - \lambda^3 c^*_3 z &   - \lambda^2 c^*_2 y &  1- \lambda^2 \hat{r}
\end{array}
\right)
\end{equation}
with:
\begin{eqnarray}
r=|a_3|^2 + |c_3|^2  \nonumber \\
r^\prime = |a_2|^2  + |b_2|^2  +|c_2|^2 \nonumber \\
\hat{r} = |c_1|^2 + |c_2|^2 + |c_3|^2 + (c_1 + c^*_1) x \\
u = a^*_2 a_3 + c^*_2 c_3  \nonumber \\
a_i = \frac{A_i}{D_i}, \quad  b_i = \frac{B_i}{D_i}, \quad
c_i = \frac{C_i}{D_i} \nonumber
\end{eqnarray}
It can be shown that a realistic $V_{CKM}$ is generated with:
\begin{equation}
|V_{cb}| \approx K_{23} \lambda^2 \qquad |V_{ub}| \approx K_{13} \lambda^3
\end{equation}
where the $ K_{ij}$ are of order one, namely $K_{23} = c_2 y$ and  $K_{13} = c_3 z$.

The general FCNC structure was analysed in detail \cite{Botella:2016ibj}
for this model, including loop-induced FCNC. The decay channels of the
vector-like quarks were also studied, together with the possibility of 
discovering them at the LHC.

\section{Conclusions} 
We emphasise a flavour fine-tuning problem present in the SM, which results 
from the 
fact that there is no reason to have a $V_{CKM}$ matrix close to the identity, 
even taking into account the strong hierarchy of quark masses. We describe a 
solution to this fine-tuning problem which involves the introduction of a 
flavour symmetry and the introduction of vector-like quarks of charges 
(-1/3) and (2/3), together with a complex singlet scalar. Prior to the 
introduction of the vector-like quarks $V_{CKM}$ equals the identity, 
but in the presence of vector-like quarks, a correct quark mass spectrum 
and a realistic CKM mixing matrix is obtained. We also point out that in
realistic models, some of the vector-like quarks are at the reach of 
second run of LHC.

\section*{Acknowledgements}
The authors thank the local organising committee of the Symposium
Discrete 2016 for the stimulating scientific atmosphere and the warm
hospitality in Warsaw, and acknowledge financial support from 
the National Science Centre, Poland, the HARMONIA project under 
contract UMO-2015/18/M/ST2/00518 (2016-2019)
to participate in the Symposium and in the 1st Harmonia Meeting that
took place immediately after the Symposium. Special thanks go to 
Maria Krawczyk.
Our work is partially supported by Funda\c c\~ ao para a Ci\^ encia
e a Tecnologia (FCT, Portugal) through the projects CERN/FIS-NUC/0010/2015,
CFTP-FCT Unit 777 (UID/FIS/00777/2013) which are partially funded through
POCTI (FEDER), COMPETE, QREN and EU.

\end{document}